\documentclass{article}

\usepackage{PRIMEarxiv}

\usepackage[utf8]{inputenc} 
\usepackage[T1]{fontenc}    
\usepackage{hyperref}       
\usepackage{url}            
\usepackage{booktabs}       
\usepackage{amsfonts}       
\usepackage{nicefrac}       
\usepackage{microtype}      
\usepackage{lipsum}
\usepackage{fancyhdr}       
\usepackage{graphicx}       
\graphicspath{{images/}}     

\usepackage{float}
\usepackage{multirow}
\usepackage[table,xcdraw]{xcolor}
\usepackage{amsmath}

\title{EndoNet: model for automatic calculation of H-score on histological slides
}

\author{
  Egor Ushakov*, Anton Naumov, Vladislav Fomberg\\
  Information Systems Department,\\
  Ivannikov Institute for System Programming of\\
  the Russian Academy of Sciences (ISP RAS)\\
  Moscow 109004, Russia\\
  \texttt{*ushakoven@yandex.ru} \\
  \And
  Polina Vishnyakova\\
  FSBI “National Medical Research Centre for\\
  Obstetrics, Gynecology and Perinatology\\
  Named after Academician V.I.Kulakov” of\\
  the Ministry of Health of the Russian Federation,\\
  Bldg 4, Oparina Street, Moscow 117997, Russia\\
  Research Institute of Molecular and Cellular Medicine,\\
  Peoples’ Friendship University of\\
  Russia (RUDN University),\\
  Miklukho-Maklaya Street 6\\
  Moscow 117198, Russia
  \And
  Aleksandra Asaturova, Alina Badlaeva, Anna Tregubova\\
  FSBI “National Medical Research Centre for\\
  Obstetrics, Gynecology and Perinatology\\
  Named after Academician V.I.Kulakov” of\\
  the Ministry of Health of the Russian Federation,\\
  Bldg 4, Oparina Street\\
  Moscow 117997, Russia
  \And
  Evgeny Karpulevich\\
  Information Systems Department,\\
  Ivannikov Institute for System Programming of\\
  the Russian Academy of Sciences (ISP RAS)\\
  Moscow 109004, Russia 
  \And
  Gennady Sukhikh\\
  FSBI “National Medical Research Centre for\\
  Obstetrics, Gynecology and Perinatology\\
  Named after Academician V.I.Kulakov” of\\
  the Ministry of Health of the Russian Federation,\\
  Bldg 4, Oparina Street\\
  Moscow 117997, Russia\\
  \And
  Timur Fatkhudinov\\
  Research Institute of Molecular and Cellular Medicine,\\
  Peoples’ Friendship University of\\
  Russia (RUDN University),\\
  Miklukho-Maklaya Street 6\\
  Moscow 117198, Russia\\
  A.P. Avtsyn Scientific Research Institute of \\
  Human Morphology\\
  Moscow 117418, Russia
}

\begin{document}
\maketitle

\begin{abstract}
H-score is a semi-quantitative method used to assess the presence and distribution of proteins in tissue samples by combining the intensity of staining and percentage of stained nuclei. It is widely used but time-consuming and can be limited in accuracy and precision. Computer-aided methods may help overcome these limitations and improve the efficiency of pathologists' workflows. In this work, we developed a model EndoNet for automatic calculation of H-score on histological slides. Our proposed method uses neural networks and consists of two main parts. The first is a detection model which predicts keypoints of centers of nuclei. The second is a H-score module which calculates the value of the H-score using mean pixel values of predicted keypoints. Our model was trained and validated on 1780 annotated tiles with a shape of  100x100 $\mu m$ and performed 0.77 mAP on a test dataset. Moreover, the model can be adjusted to a specific specialist or whole laboratory to reproduce the manner of calculating the H-score. Thus, EndoNet is effective and robust in the analysis of histology slides, which can improve and significantly accelerate the work of pathologists.
\end{abstract}


\section{Introduction}
Immunohistochemical analysis is a classic technology used to assess the expression and spatial distribution of a particular protein biomarker in tissue samples. This method is widely used in clinical practice, especially in diagnostics in oncology such as identification and classification of a tumor, its localization, mutation-specific status. Qualitative and semi-quantitative evaluation of slides is carried out by a pathologist, which has a number of limitations including, for example the subjectivity of the specialist's evaluation, the limited range of grades of staining (scale from 0 to 3+), the time-consuming nature of the assessment, the complexity of scaling the process for large projects. An increasing number of works indicate that computer-aided analysis can replace or at least facilitate and standardize the work of a pathologist \cite{Rizzardi2012, Srinidhi_2021, Iizuka2020}. H-score (H-score), a method of assessing the extent of nuclear immunoreactivity, applicable to steroid receptors, is used as a semi-quantitative assessment of the degree of slide staining, first mentioned in the works of the 80s \cite{Budwit_Novotny1986, vanNetten1987}. Since then, H-score has been successfully applied in immunohistochemical analysis\cite{Babu2020, Pierceall2011, Sharada2021} and has recently even been reborn into a digital image analysis-based approach, called the pixelwise H-score\cite{Ram2021}. Automated H-score proves to be a successful and efficient alternative to visual H-score.

With the invention of the first whole-slide scanners in the 1990s \cite{Pantanowitz2018} it became possible to make whole slide images of various tissues with a resolution at the level where nuclei become distinct. With the advent of the first digital slides, the first algorithms for their analysis appeared, which included trainable and non-trainable algorithms \cite{Gurcan2009, Madabhushi2016, Veta2014, Li2020}. Whole slide image analysis has many challenging aspects such as large slide sizes in pixels while the region of interest may be less than 10 pixels. Also a slide could contain a large number of such regions in different places, the information about which needs to be aggregated and analyzed. Therefore, the whole slide is often cut into small tiles and processed separately, with subsequent aggregation of the information received from each tile.

Digital pathology tools, such as neural networks, can significantly facilitate the work of pathologists in that they provide results with accuracy and precision, that conventional methods cannot provide. Whole slide images contain a large amount of information that is not always visible to the human eye, especially when it is necessary to process myriads of similar slides in a row. Therefore, pathologists can process only a small part of the entire image, the so-called "region of interest", which can bias the final score of the whole slide. These factors contributed to the growth of interest in the use of neural networks for the analysis of whole slide images, in particular, histology images. Neural networks rose to prominence with the invention of AlexNet \cite{alexnet} and later, when neural networks outperformed humans in the classification of images \cite{He_2015_ICCV}, which further increased interest in the development of deep learning methods. Later, convolutional neural networks (CNN) were invented, which proved themselves to be effective in image analysis and processing. At the moment, convolutional neural networks are quite popular in the analysis of histology images such as detection, tissue classification, annotation, quantification. Their high performance means that they do not require a large amount of training data. This contrasts with visual transformers \cite{visual_transformer} which are only gaining popularity. Therefore, in this paper, a convolutional architecture was chosen for our model.

There are many different variations of convolutional neural network architectures, such as VGG, ResNet, Unet and etc. For medical images, the Unet architecture was specially invented, which has gained great popularity in this field. It looks like an HourGlass, compressing and decompressing the image, converting the input image into a heatmap.

To calculate the H-score, we need to know the number of nuclei of different levels of staining on the slide. Neural networks can perform as a detector that will find cells on the slide and classify them, which will allow us to calculate the H-score.

Using all these approaches, we present EndoNet, the model for automatic calculation H-score on histological slides.


\section{Methods}
EndoNet is a model for predicting H-score in the stroma and epithelium on endometrium slides. The model consists of two important parts. The first part is a detection model that detects nuclei and predicts their keypoints by input tiles. The second part is a module for calculating the H-score based on the predicted keypoints.

This section will describe the data on which the model was trained and validated, where they were obtained from, and how exactly the tiles were obtained from whole slide images.

There will be a description of the detection model and its architecture, how the training was performed, and the choice of optimal parameters for it.

Furthermore, we will also describe what the H-score calculation module consists of, what calculation formula it uses, and how the level of staining is determined.

\subsection{Data}

\begin{figure}[H]
    \centering
    \includegraphics[width=13cm]{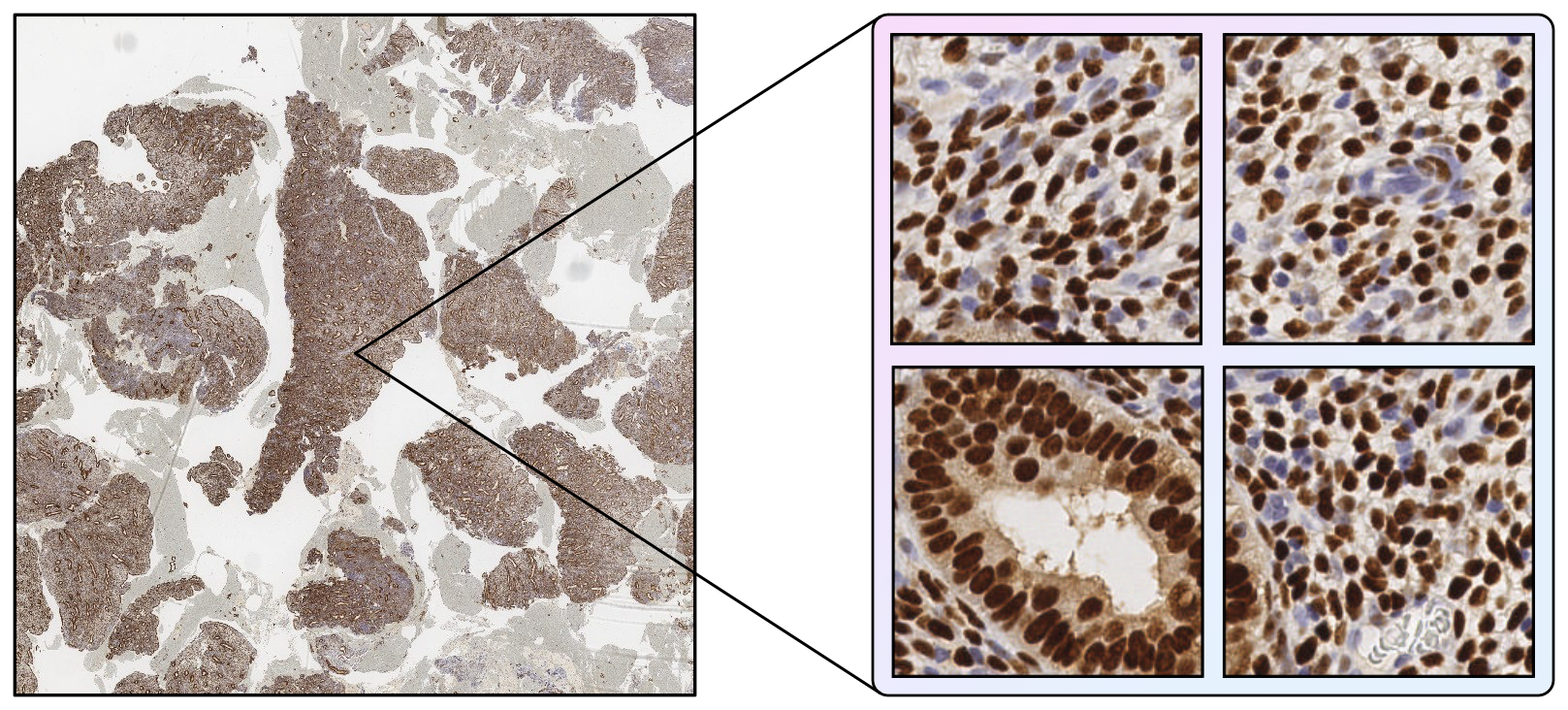}
    \caption{Example of whole slide (left) and cutted tiles (right).}
    \label{fig:tiles}
\end{figure}

Our dataset consists of slides of endometrium after Immunohistochemical (IHC) analysis. The slides were taken from two different sources: EndoNuke\cite{endonuke} open histology dataset (the "bulk" part) and a pathology laboratory dataset (hereinafter called PathLab dataset). There are some differences in the slides, such as methods of staining, equipment, reagents for staining, types of tissue etc., which make the detection a challenging task. Whole slides were cut into small tiles, examples are shown in Figure \ref{fig:tiles}. Tiles in EndoNuke are images of various sizes in pixels (mainly 200x200 and 400x400 pixels), because slides have different $\mu m/pixel$ values. But they capture the same field of view of 100x100 $\mu m$. All tiles were resized into 512x512 pixels before being fed as input to the model. Tiles from the pathology laboratory are primarily 395x395 pixels, but they also capture the field of view of 100x100 $\mu m$. Slides from PathLab were received without annotation, thus these tiles were annotated. The protocol for annotation tiles from the laboratory matches the EndoNuke\cite{endonuke} annotation protocol. Overall, 40 tiles were annotated.

All datasets were merged and split into training, validation and test parts in proportion of 3:1:1. The training part was used to train the model, the validation part was used to select the best epoch during the training, and the test part was used to evaluate the model.


\subsection{Detection model architecture}
The main concept in the Detection is how to represent the position of an object. This can be represented as a Bounding Box (a vector of length of 4, which contains the coordinates of the upper-left and lower-right) and the coordinates of the center of the object (a vector of length of 2). 

There are many methods and models that use approach with Bounding Boxes. For instance, in \cite{Sun2021} the authors used convolutional neural networks, where ResNet\cite{resnet} and ResNeXt\cite{resnext} were used as the backbone. The main feature of this work is the presence of additional layers with embeddings that improve the detection and classification of nuclei, unlike standard models of this type, such as \cite{faster_rcnn}.

Also, there is another approach to detection, which we consider more native to our task. Since we only need the 2 coordinates of the nuclei centers instead of 4 for bounding boxes, we reduce the number of predicted values by two times. Instead of directly predicting the coordinates of objects, it is possible to predict a probability map, called a heatmap. A heatmap is a field of probabilities of the object being at each point, and the higher the value in the pixel, the greater the probability of finding the object. Accordingly, the maximum of the probability peak will correspond to the center of the cell nucleus. The coordinates of the peaks on the heatmap, which can be found with the max-pooling procedure, will be the coordinates of the center of the expected object.

Predicting keypoints using heatmaps is an image-to-image task, and convolutional neural networks perform well in it. After getting heatmaps as logits of neural networks we need to extract keypoints from them. Heatmaps go through a max-pooling procedure to get coordinates and classes of local maximums and form keypoints. Following that, the keypoints can be post-processed which includes thresholding by  the confidence level and a Weighted boxes fusion procedure\cite{WBF}.

\begin{figure}[H]
    \centering
    \includegraphics[width=13cm]{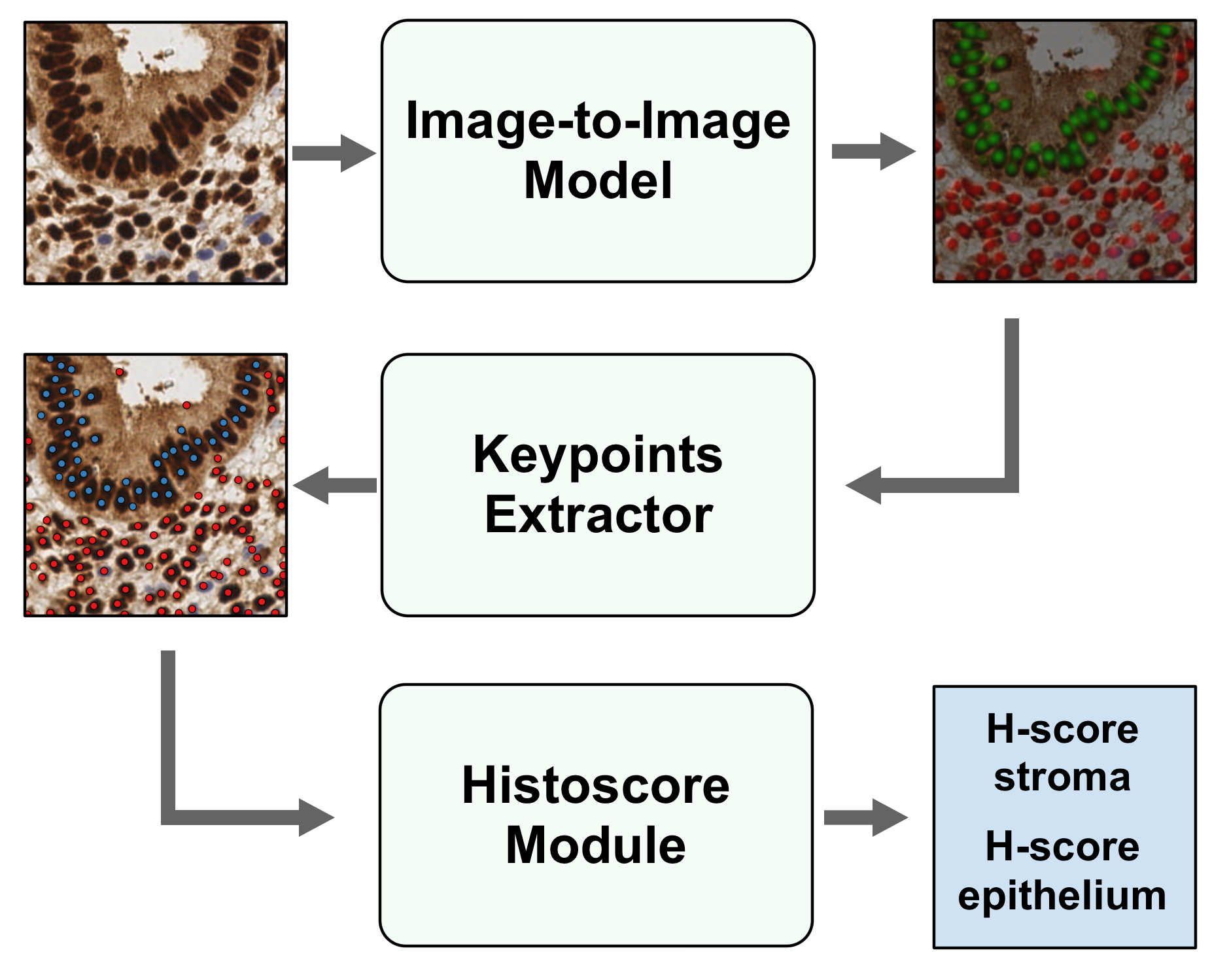}
    \caption{Architecture of EndoNet model. Tiles go through Image-to-Image model to be converted into heatmaps, Keypoint Extractor gets coordinates and classes of the centers of nuclei and passes them to H-score Model to calculate H-score in stroma and epithelium.}
    \label{fig:model}
\end{figure}

Summing all these ideas, the detector in EndoNet consists of Image-to-Image neural network to predict heatmaps of nuclei and the Keypoint Extractor to get keypoints from heatmaps. Its architecture is illustrated in Figure \ref{fig:model}.

As Image-to-Image model we used such architectures as UNet, UNet++, U2Net, FPNet and others. These models consist of an encoder, which compresses and extracts essential information from the image, and a decoder, which expands the image after the encoder. The architecture of the decoder is mirrored to the architecture of the encoder.

\subsection{Training of detection model}

To find the optimal architecture for the detection model, a grid search of parameters was performed. The set of optimized parameters included: the backbone architecture, the parameters of the Keypoint Extractor (threshold, minimum distance between keypoints, the value of pooling). The optimized value was the mean Average Precision (mAP). The optimization step consists of training the model during 100 epochs with the current set parameters described above, choosing the best epoch based on the performance metric on a validation dataset and saving the results. 100 epochs were chosen because our models are mainly trained to a plateau in 40-50 epochs, but a surplus was made for architectures that may need more epochs.

After doing a grid search and choosing the optimal architecture of the detection model, such parameters as schedulers, optimizers and augmentations were selected separately from each other.

From the list of loss functions: Huber loss, MSE, Gaussian Focal loss (similar to a common Focal loss\cite{focal} function but for a continuous distribution), we chose a Huber loss function because it performed effective in our previous tasks:

\begin{equation}
L_{huber}=
\begin{cases}
\frac{1}{2}(y - \hat{y})^{2} & if \left | (y - \hat{y})  \right | < \delta\\
\delta ((y - \hat{y}) - \frac1 2 \delta) & otherwise
\end{cases}
\end{equation}

\subsection{Pre-training}

To improve the generalization ability of the our CNN model, self-supervised learning techniques were proposed.
Pre-training means training a model on a larger dataset before training a similar model on a similar dataset. A scheme of the pre-training pipeline is shown in Figure \ref{fig:pretraining}:

\begin{figure}[H]
    \centering
    \includegraphics[width=13cm]{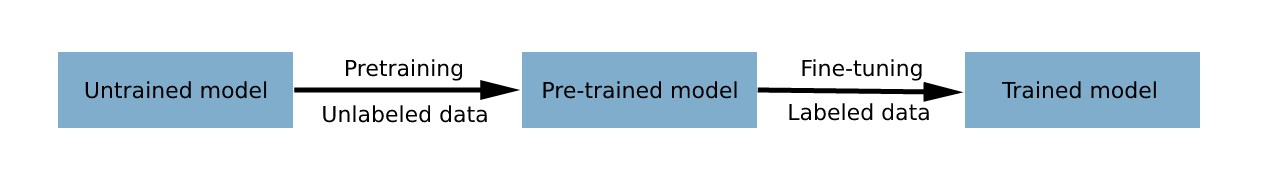}
    \caption{General pipeline of pre-training process.}
    \label{fig:pretraining}
\end{figure}

\begin{figure}[H]
    \centering
    \includegraphics[width=8cm]{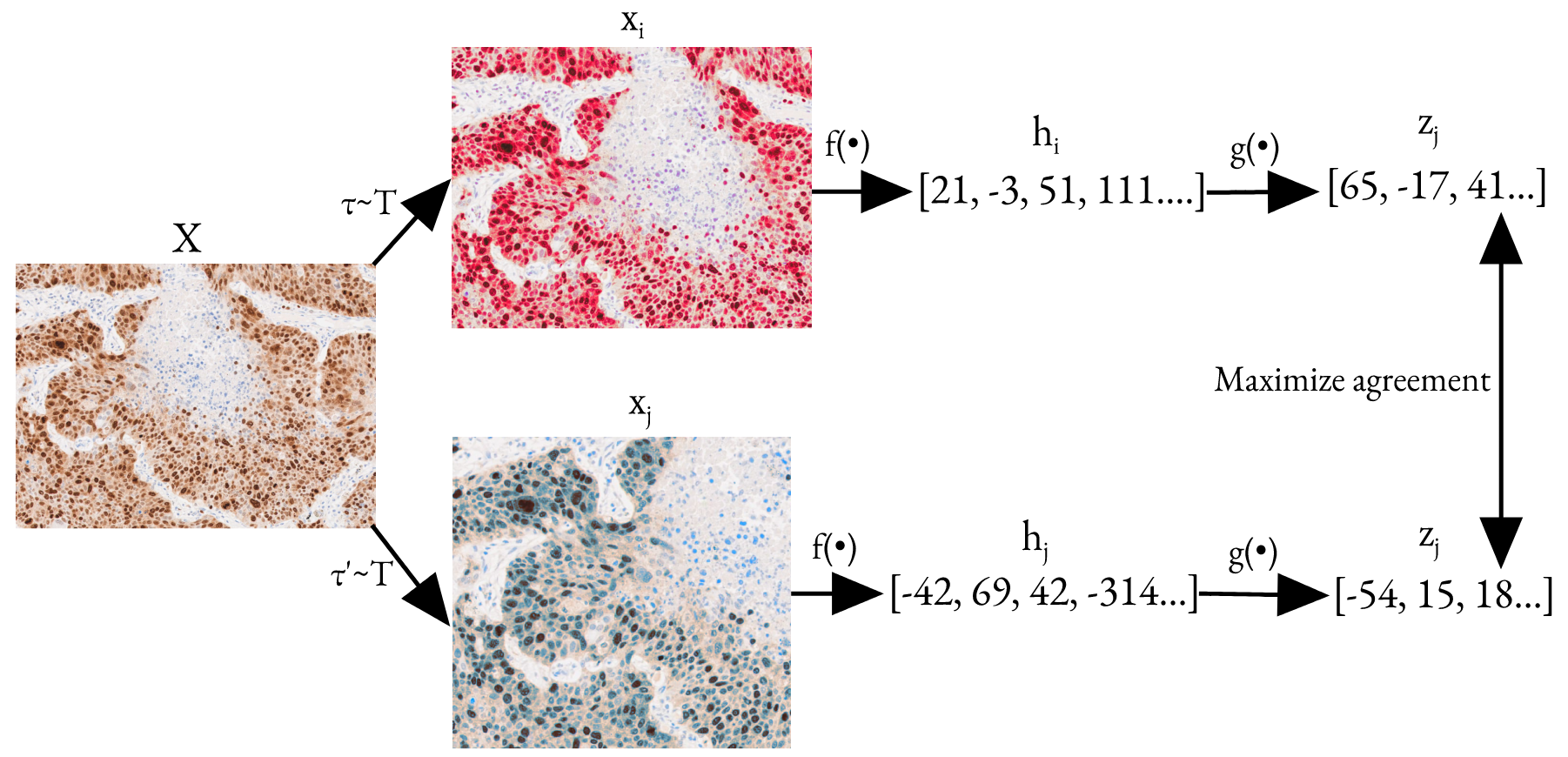}
    \caption{Pre-training process with SimCLR\cite{SimCLR}. Here $t \in \tau$ and $t' \in \tau$ are two augmentations taken from the same family of augmentations. $f(\cdot)$ is a base encoding network and $g(\cdot)$ is a projection head that maps hidden representation to another space, where contrastive loss is applied. X is initial image, $x_i$ and $x_j$ are augmented images, $h_i$ and $h_j$ are hidden representations of corresponding augmented images, and $z_i$ and $z_j$ are output of decoding network. Optimization task here is to maximize agreement between $z_j$ and $z_i$.}
    \label{fig:simclr}
\end{figure}

The generally accepted approach is to initialize neural network with weights obtained as a result of pre-training on the ImageNet\cite{ImageNet} dataset. However, the ImageNet dataset differs greatly from the EndoNuke and PathLab datasets. The proposed solution is to use unlabeled images with the SimCLR \cite{SimCLR} method to improve the generalization ability of the model without increasing the number of labeled images. The SimCLR learns by maximizing the level of agreement between differently augmented views of the same image with contrastive loss. The augmentations consist of random cropping followed by resizing back to the original size, random color distortion and random Gaussian blur. As stated in the original paper \cite{SimCLR}, this method benefits from larger batch sizes. The Figure \ref{fig:simclr} describes how exactly the model is trained with the SimCLR model.

The unlabeled part of the Pathlib dataset slides was filtered to exclude empty images by comparing the mean and standard deviation of choosing images with the corresponding threshold values. Cutting slides into images is the same as in the labeled part. The resulting dataset contains 877 286 unlabeled images. We decided to test a model with an architecture that performed best in a grid search. A ResNet50 encoder with random initialized weights was pre-trained on an unlabeled dataset with SimCLR method for 80 epochs. A model with the SimCLR pre-trained weights and a model with ImageNet pre-trained weights were compared against each other on a test dataset by mAP metric. Metric was calculated for each batch, to get a set of results instead of one number. The results are shown in the table\ref{table:1}:

\begin{table}
\centering
\begin{tabular}{lll}
\hline \hline
                       & \textbf{SimCLR pre-trained} & \textbf{ImageNET}\\ \hline
\textbf{Stroma AP}     & 0.8577                & 0.8544\\ 
\textbf{Epithelium AP} & 0.7576                & 0.7256\\ 
\textbf{mAP}           & 0.8077                & 0.7900\\ \hline \hline
\end{tabular}
\caption{Resulting metrics, computed on test dataset}
\label{table:1}
\end{table}

To draw more meaningful conclusions, predictions of the baseline model were subtracted from predictions of the pre-trained model. The resulting distribution was resampled with the bootstrap method, which implies random sampling with replacement, 10 000 resamples in total. Then confidence intervals were calculated with a confidence level of 0.95. The bootstrapping procedure was repeated 10 000 times, and the results were averaged. It is necessary for obtaining objective results, because bootstrapping is non-deterministic. The results are shown in the table\ref{table:2}:

\begin{table}
\centering
\begin{tabular}{lll}
\hline \hline
                       & \textbf{Lower bound} & \textbf{Upper bound} \\ \hline
\textbf{Stroma AP}     & -0.00666            & 0.01840             \\ 
\textbf{Epithelium AP} &  0.00461            & 0.07010            \\ 
\textbf{mAP}           & -0.00024            & 0.04115          \\ \hline \hline
\end{tabular}
\caption{Confidence interval for mean difference in models }
\label{table:2}
\end{table}


\subsection{H-score module}
To calculate H-score we should add up weighted counts of no, weak, moderate and strong stained nuclei in the stroma and epithelium:

\begin{equation}
    H{-}score = 0 * none + 1 * weak + 2 * moderate + 3 * strong
\end{equation}

The EndoNet determines the degree of staining of each predicted by the detection model nuclei. It takes some area around the predicted keypoint (a square slightly smaller than the average radius of the core calculated in EndoNuke) and averaged over each of the three channels. As a result, a vector of length 3 was obtained for each keypoint.

There are two types of nuclei on the tiles: unstained and stained. Unstained cells are blue cells that should not be counted. Stained cells are brown, have three degrees of staining: weak, moderate and strong stained. The task of determining the degree of staining was divided into two subtasks: separation of unstained from stained nuclei and further classification of stained nuclei.

There are some works in this field which provide methods for nuclei with different staining on histology slides \cite{Rizzardi2012, Krajewska2009} (usually blue nuclei from brown). Many of them use specific transformations of the RBG-space of input images. The most common transformation is a transformation into HSV-space, where Red, Blue and Green channels transform into Hue, Saturation and Value. We also use this transformation.

\begin{figure}[H]
    \centering
    \includegraphics[width=13cm]{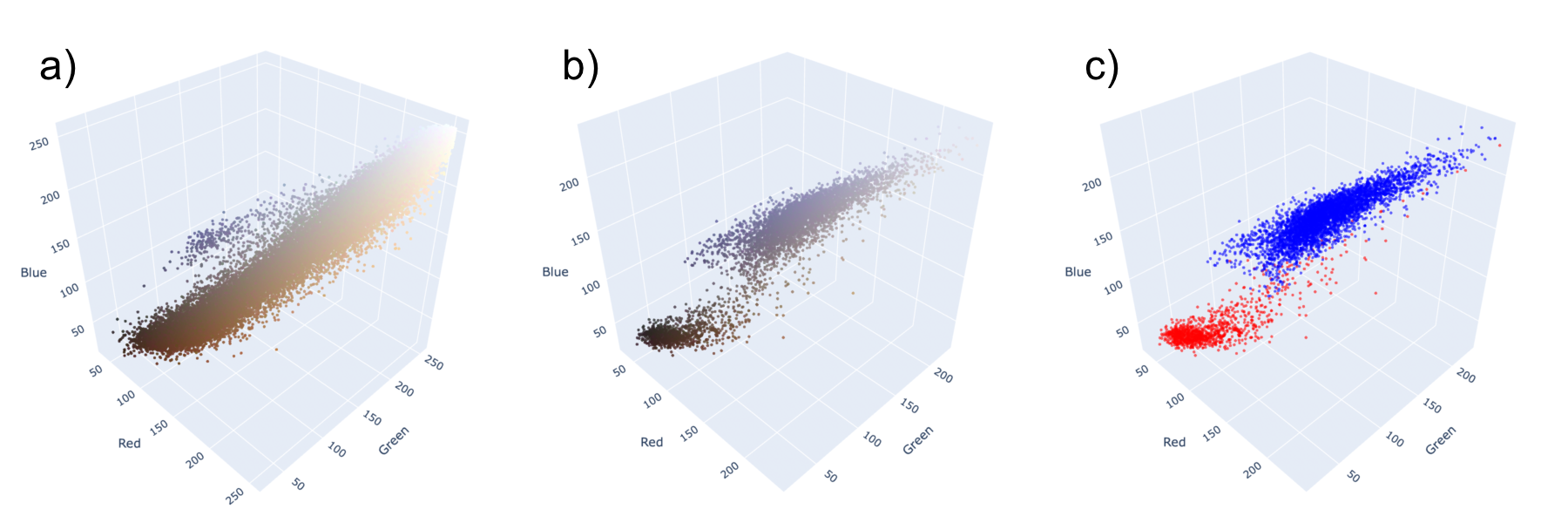}
    \caption{Distributions of pixels of a) the whole tile, b) blue and brown nuclei, stained in their colors, c) blue and brown nuclei, where red are brown nuclei, blue are blue nuclei.}
    \label{fig:distribution_1}
\end{figure}

\begin{figure}[H]
    \centering
    \includegraphics[width=13cm]{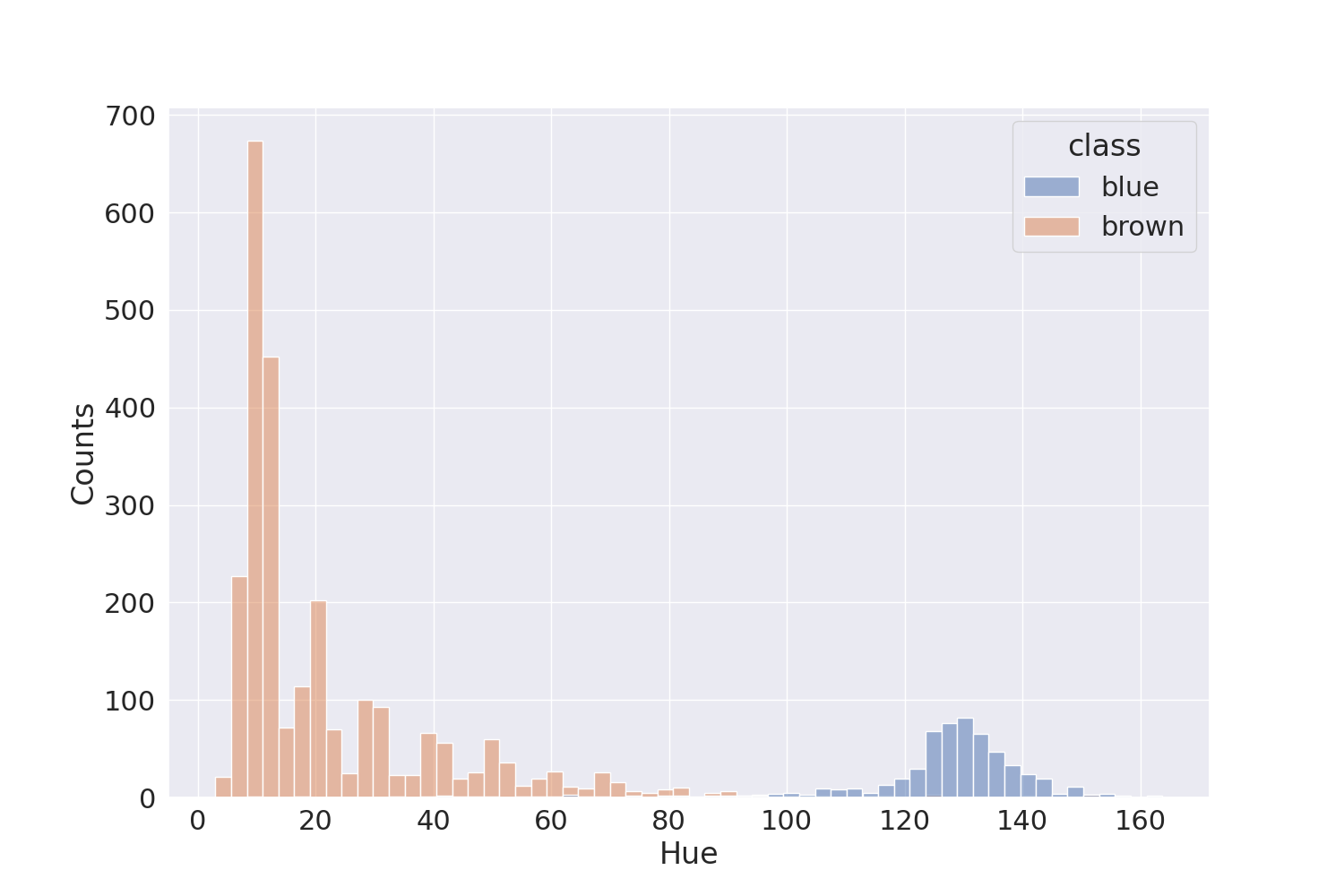}
    \caption{Distribution of blue and brown nuclei in Hue channel.}
    \label{fig:distribution_3}
\end{figure}

The first task is to separate blue and brown nuclei on slides. As we see in Figure \ref{fig:distribution_1}, blue and brown nuclei are pretty well separated even in RGB-space. The distribution shown in the picture is based on the annotation by pathologists. The HSV-space is useful too in this case, that blue and brown color are far from each other on Hue axis as shown in Figure \ref{fig:distribution_3}. Thus, it is easily to separate most of them (except nuclei with very weak staining). Since the two distributions are far apart, the threshold can be taken as the average between the two peaks of the distributions.

The second task is to classify stained brown nuclei to 3 classes: weak, moderate and strong. However, there are no solid thresholds for these classes, and classification is a very subjective process. So, we annotated 40 tiles by two pathologists for eight classes: no, weak, moderate and strong stained nuclei for stroma and epithelium. Pathologists annotated tiles separately from each other.

\begin{table}
\centering
\begin{tabular}{llll}
\hline \hline
                       & \textbf{EndoNuke} & \textbf{PathLab} & \textbf{Combined} \\ \hline
\textbf{Stroma AP}     & 0.85               & 0.83                & 0.85         \\ 
\textbf{Epithelium AP} & 0.69               & 0.84                & 0.69         \\ 
\textbf{mAP}           & 0.77               & 0.84                & 0.77         \\ \hline \hline
\end{tabular}
\caption{Results computed on test sample}
\label{table:3}
\end{table}

\begin{figure}[H]
    \centering
    \includegraphics[width=13cm]{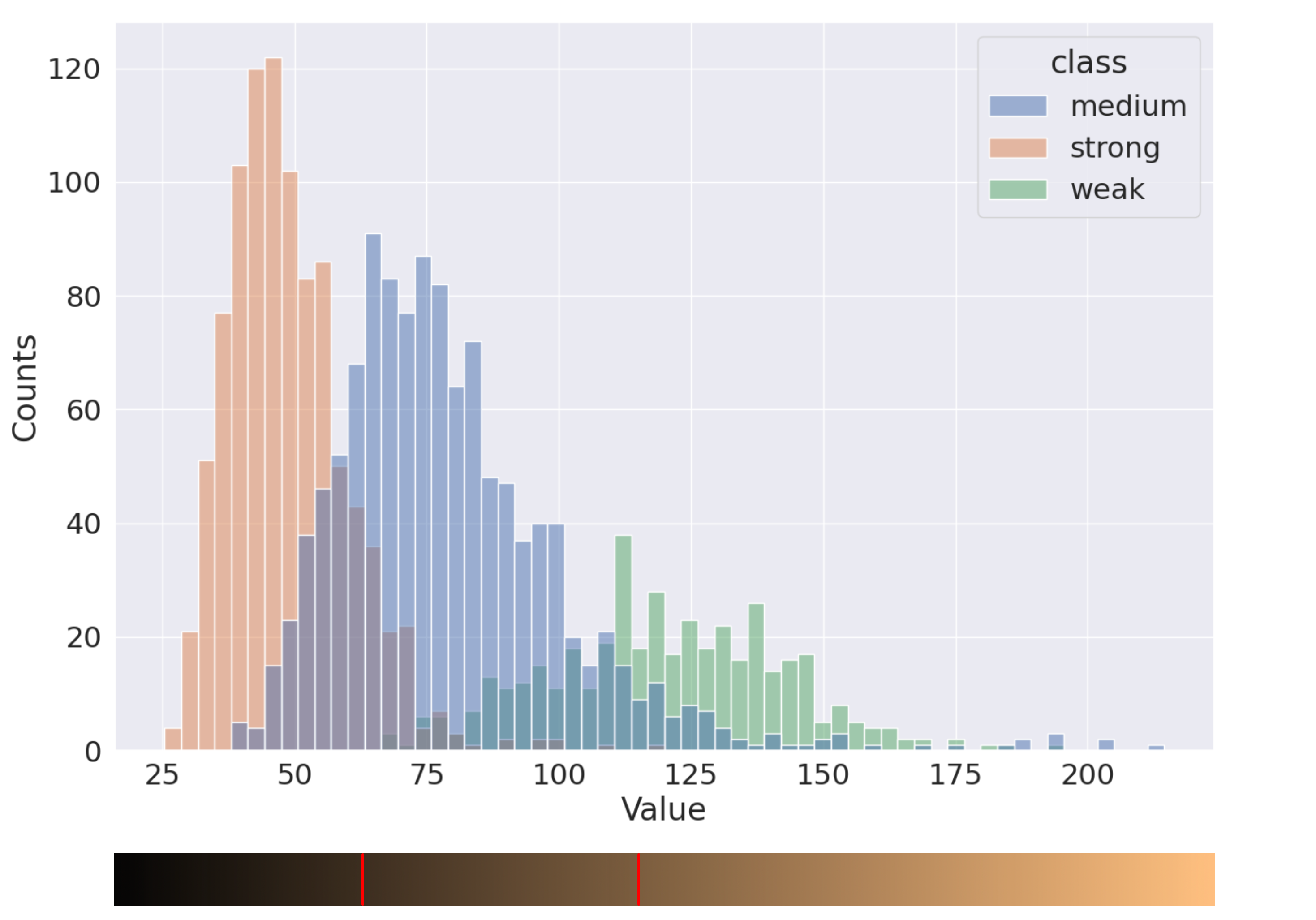}
    \caption{Distribution of brown nuclei in Value channel.}
    \label{fig:distribution_2}
\end{figure}

The distribution of pixels into three classes, marked by histopathologists, is presented in Figure \ref{fig:distribution_2}. The distributions of weak and strong nuclei are easily distinguishable but the moderate intersects a lot with others. With this distribution it is possible to separate nuclei to three classes by two thresholds. To determine the value of these thresholds, we iterated over a set of possible parameters and calculated the deviation between two H-scores: H-score based on annotation made by pathologists and H-score, based on model predictions and the current set of thresholds. For instance, we got values 60 and 105 of thresholds for the distribution in Figure \ref{fig:distribution_2}.

Using an algorithm for selecting parameters, you can create a profile with thresholds for each pathologist. To do this, you first need to calibrate the model: several different regions of interest are selected and annotated. The model selects thresholds for annotated regions of interest. The received thresholds are saved for each doctor and can be used again.


\section{Results}
\subsection{Training}

After finishing a grid search we got "a best" model with a UnetPlusPlus with a ResNet50 backbone, because it achieved the highest score. Moreover, during all iterations we used image augmentations such as rotations, flips, Gauss noise, HSV shift and others. As a quality metric we use mAP. The mAP metric is used for object detection tasks. Traditionally, predictions of object detection models are presented as BBoxes whereas our model predictions are presented as Keypoints. To determine TP, FP, FN, we use Minkowski distance with the degree two (which is Euclidean distance). As a threshold distance, we use the value of the average radius of the nuclei cell obtained in EndoNuke\cite{endonuke}. The training was conducted on a combined dataset and results are presented in Table \ref{table:3}. The PathLab makes up a small part of the combined test dataset, so it does not contribute much to the overall result.

\begin{table}
\centering
\begin{tabular}{llll}
\hline \hline
\textbf{Annonator} & \textbf{Slide} & \textbf{Left} & \textbf{Right} \\ \hline
\multirow{4}{*}{1st} & 1              & 80            & 120            \\ 
                   & 2              & 80            & 125            \\ 
                   & 3              & 80            & 120            \\ 
                   &  4              &  80            &  125            \\ \hline
\multirow{4}{*}{2nd} &  4              &  80            &  135            \\ 
                   & 5              & 80            & 120            \\ 
                   & 6              & 75            & 130            \\ 
                   & 7              & 80            & 130            \\ \hline \hline
\end{tabular}
\caption{Calculated thresholds of "Value" dimension in HSV space for each slide for both annotators. The "Left" means threshold which divides strong and moderate staining, the "Right" means threshold which divides moderate and weak staining such in Figure \ref{fig:distribution_2}. The fourth slide is mutual for calculating the agreement level.}
\label{table:4}
\end{table}

\subsection{H-score}
We annotated seven whole slide images (Progesterone and Estrogen tiles) with five tiles per each slide. The fourth slide was mutual for both pathologists to measure the level of agreement between them. Then we calculated thresholds for dividing colors of nuclei in the following way: we used three slides as a training dataset to find optimal thresholds for the fourth, and we did so for each slide. For instance, to calculate thresholds for the third slide, we use first, second and fourth slides in the training process. The results are presented in Table \ref{table:4}.

\begin{table}
\centering
\begin{tabular}{lllll}
\hline \hline
\textbf{Annonator}    & \multicolumn{1}{c}{\textbf{\begin{tabular}[c]{@{}c@{}}Slide \\ number\end{tabular}}} & \multicolumn{1}{c}{\textbf{Manual}} & \textbf{\begin{tabular}[c]{@{}l@{}}Model\\ small\end{tabular}} & \multicolumn{1}{c}{\textbf{\begin{tabular}[c]{@{}c@{}}Model\\ big\end{tabular}}} \\ \hline
                      & 1                                                                                     & 137                                  & 120                                                            & 122                                                                               \\ 
                      & 2                                                                                     & 149                                  & 165                                                            & 164                                                                               \\ 
                      & 3                                                                                     & 138                                  & 128                                                            & 116                                                                               \\ 
\multirow{-4}{*}{1st} &  4                                                             &  183          &  178                                    &  179                                                       \\ \hline
                      &  4                                                             &  187          &  181                                    &  184                                                       \\ 
                      & 5                                                                                     & 131                                  & 109                                                            & 114                                                                               \\ 
                      & 6                                                                                     & 198                                  & 165                                                            & 158                                                                               \\ 
\multirow{-4}{*}{2nd} & 7                                                                                     & 180                                  & 168                                                            & 188                                                                               \\ \hline \hline
\end{tabular}
\caption{Calculated H-score in stroma for each slide for both annotators. These scores for each slide are calculated based on thresholds from Table \ref{table:4}. The "Manual" H-score is based on keypoint annotation of pathologists, the "Model small" H-score is based on annotation provided by our model on the same tiles, the "Model big" H-score is based on annotation provided by our model but on the big amount of tiles from the same slides.}
\label{table:5}
\end{table}

\begin{table}
\centering
\begin{tabular}{lllll}
\hline \hline
\textbf{Annonator}    & \multicolumn{1}{c}{\textbf{\begin{tabular}[c]{@{}c@{}}Slide \\ number\end{tabular}}} & \multicolumn{1}{c}{\textbf{Manual}} & \textbf{\begin{tabular}[c]{@{}l@{}}Model\\ small\end{tabular}} & \multicolumn{1}{c}{\textbf{\begin{tabular}[c]{@{}c@{}}Model\\ big\end{tabular}}} \\ \hline
                      & 1                                                                                     & 164                                  & 145                                                            & 161                                                                               \\ 
                      & 2                                                                                     & 180                                  & 182                                                            & 181                                                                               \\ 
                      & 3                                                                                     & 144                                  & 144                                                            & 128                                                                               \\ 
\multirow{-4}{*}{1st} &  4                                                             &  137          &  159                                    &  141                                                       \\ \hline
                      &  4                                                             &  150          &  167                                    &  159                                                       \\ 
                      & 5                                                                                     & 150                                  & 142                                                            & 138                                                                               \\ 
                      & 6                                                                                     & 57                                   & 88                                                             & 65                                                                                \\ 
\multirow{-4}{*}{2nd} & 7                                                                                     & 202                                  & 198                                                            & 219                                                                               \\ \hline \hline
\end{tabular}
\caption{Calculated H-score in epithelium for each slide for both annotators. These scores for each slide are calculated based on thresholds from Table \ref{table:4}. The "Manual" H-score is based on keypoint annotation of pathologists, the "Model small" H-score is based on annotation provided by our model on the same tiles, the "Model big" H-score is based on annotation provided by the our model but on the big amount of tiles from the same slides.}
\label{table:6}
\end{table}

After calculating the thresholds, we measured the H-score for each slide using these thresholds. Our Neural Network provides keypoint annotations: coordinates of centers and classes of nuclei. To classify stained nuclei into three classes: strong, moderate and weak, we transformed RGB image into HSV, took pixels in a small area around keypoints, calculated the mean value of the "Value" channel and compared with the thresholds. The results are presented for stroma and epithelium in Table \ref{table:5} and Table \ref{table:6} respectively. Model H-scores differ from manual ones, but such deviations are normal. Many laboratories could differ from each other by even greater values. Model H-scores of the fourth slide are similar that means EndoNet is stable.

\begin{figure}[H]
    \centering
    \includegraphics[width=13cm]{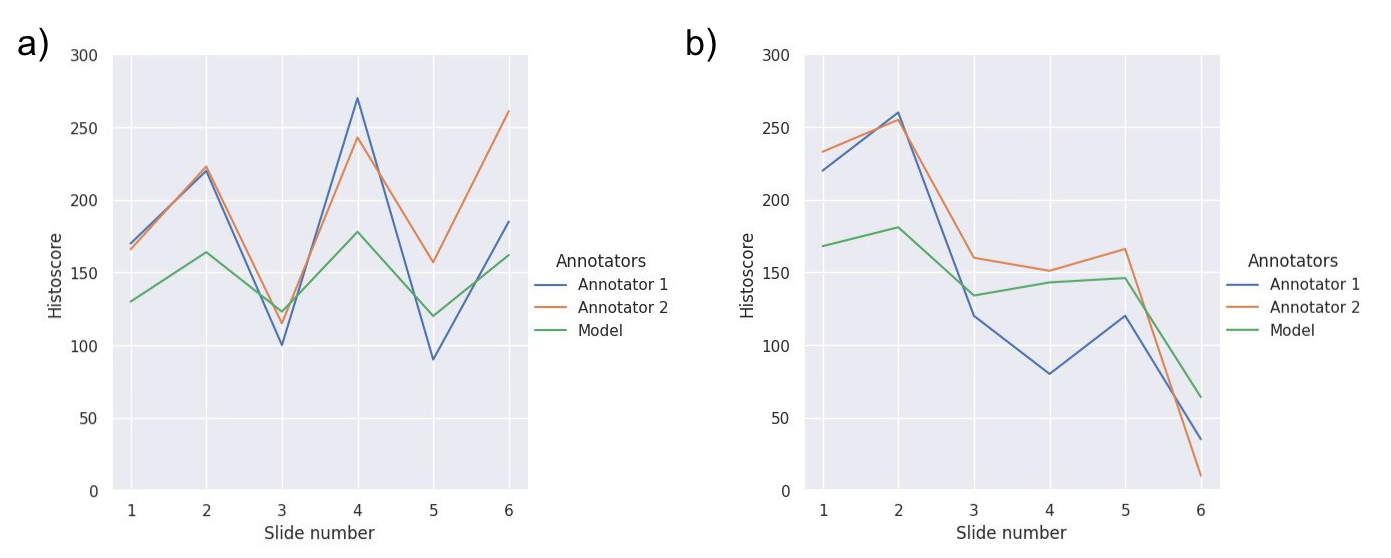}
    \caption{H-scores by pathologists and EndoNet model for 6 slides in a) stroma and b) epithelium.}
    \label{fig:H-scores}
\end{figure}

Furthermore, an additional comparison was conducted to assess the H-score results by pathologists and the model. The H-score assessment by pathologists was performed in a "working conditions" setting, wherein cells were not marked in selected regions of interest and a formula(2) for counting was not used. Instead, pathologists relied on their internal intuition and approximate estimation of H-score. The results are illustrated in Figure \ref{fig:H-scores}. The model utilized the average threshold values provided in Table \ref{table:4} to evaluate the H-score. 

\section{Discussion}

Immunohistochemical (IHC) analysis plays a crucial role in assessing protein biomarker expression in tissue samples, especially in oncology diagnostics. By visualizing the presence and distribution of specific proteins in tissues, IHC analysis provides valuable insights into disease diagnosis, prognosis, and treatment selection. However, traditional IHC analysis has limitations that can impact its reliability and reproducibility. These limitations include subjectivity in interpretation, a limited grading scale for scoring, and time-consuming manual evaluation, which can lead to inter-observer variability and inconsistency in results.

To overcome these challenges, computer-aided analysis, including automated H-score, has emerged as a promising alternative to standardize and streamline the IHC analysis process. The utilization of computer-aided analysis, has the potential to greatly enhance the efficiency and accuracy of IHC analysis in clinical practice, leading to more reliable and reproducible results in many fields of diagnostic.

In this paper, we propose EndoNet, a CNN-based model designed to automatically calculate H-score on histological slides. EndoNet can accurately and objectively analyze IHC-stained slides, reducing subjectivity, and significantly speeding up the evaluation process. It achieved high results such as 0.77 mean Average Precision on a test dataset. Furthermore, the model can be customized to suit the preferences of a specific specialist or an entire laboratory, enabling the replication of their preferred style of calculating the H-score.

In recent years, the application of neural networks in the analysis of histological images has emerged as a promising approach for automated and efficient image analysis in various medical and scientific domains. Currently, several other models used for analyzing IHC slides are developed. Hao Sun et el.\cite{Sun2020} developed a CAD approach based on a convolutional neural network (CNN) and attention mechanisms. They used the ten-fold cross-validation on ~3,300 hematoxylin and eosin (H\&E) image patches from ~500 endometrial specimens and outperformed three human experts and five CNN-based classifiers regarding overall classification performance.
Amal Lahiani et al.\cite{Lahiani2019} presented a deep learning method to automatically segment digitised slide images with multiple stainings into compartments of tumour, healthy tissue, necrosis, and background. Their method utilizes a fully convolutional neural network (CNN) that incorporates a color deconvolution segment, trained end-to-end. This color deconvolution segment aids in accelerating the network's convergence and enables it to effectively handle staining variability within the dataset. Also they used 77 whole slide images of colorectal carcinoma metastases in liver tissue from biopsy slides stained with H\&E (blue, pink) and 8 additional immunohistochemistry (IHC) slides.
Harshita Sharma et al. \cite{Sharma2017} analyzed H\&E whole slide images of gastric carcinoma with the help of deep learning methods in digital histopathology. Their proposed convolutional neural network architecture reports classification accuracy of 0.6990 for cancer classification and 0.8144 for necrosis detection.
A technique for automatic immune cell counting on digitally scanned images of immunohistochemistry (IHC) stained slides was presented by Ting Chen et al\cite{Chen2014}. The method employs a sparse color unmixing approach to segregate the IHC image into distinct color channels that correspond to different cell structures. The algorithm's performance was evaluated on a clinical dataset that comprised a substantial number of IHC slides.

Despite the studies being most closely aligned with our task, they are trained on data from a distinct domain, encompassing different tissue types and coloring methods. Furthermore, these studies lack a crucial H-score evaluation module that is of significance to our research. In addition, our model is unique in task of assessing endometrium receptivity. Furthermore, our model offers the flexibility of individual customization for pathologists or the entire laboratory, rendering it a versatile tool in addressing specific requirements.

The H-score method for assessing the presence and distribution of proteins in tissue samples has some limitations as it is time-consuming and lacks accuracy and precision. We propose a solution to these issues by developing a computer-aided method called EndoNet, which uses neural networks to automatically calculate the H-score on histological slides.

EndoNet consists of two main parts: a detection model that predicts keypoints of centers of nuclei, and a H-score module that calculates the value of the H-score using mean pixel values of predicted keypoints. The model was trained and validated on a set of annotated tiles and achieved high score on a test dataset. Additionally, the model can be customized for specific specialists or laboratories to reproduce the manner of calculating the H-score.

The development of EndoNet can have significant benefits for pathologists, as it could improve the accuracy and efficiency of H-score calculations, which are important for the diagnosis and treatment of many diseases. However, EndoNet has some limitations.

The first is that the detection model proves to be less accurate on tiles from other labs. To mitigate this constraint and to increase the generalizing ability of the model, several techniques can be explored, such as expanding the training dataset with slides from other labs or using special augmentation.

In addition, we carried out some experiments involving augmentations aimed at increasing the metric value of a model trained on tiles from one lab and tested on tiles from another lab. Notably, the "HSV shift" augmentation yielded a significant improvement in the quality of the results. We maintain that further research on augmentations that are even more effective could lead to further enhancements in the quality of the model.

The second limitation pertains to the disparity between the real data and the data present in the dataset. To evaluate the H-score, a specialist is required to manually annotate and assess the color of each nucleus in the selected areas of interest. This is a time-consuming and monotonous task, particularly when evaluating numerous slides. Consequently, pathologists infrequently utilize this method to evaluate H-score. Instead of counting the number of weak, moderate, and strong nuclei, pathologists estimate the ratio of stained nuclei approximately. This H-score assessment method is commonly employed in practice because of the large volume of samples that need to be processed each day. As a result, such results often vary from those obtained with an accurate assessment of stained nuclei. As depicted in the Figure \ref{fig:H-scores}, the results differ significantly, and pathologists using this method often overestimate the H-score value. However, pathologists and the model reproduce each other's H-score line shape.

It is our belief that, upon addressing all the limitations, the model will be capable of demonstrating even more impressive outcomes. Nonetheless, EndoNet has already exhibited promising results and is endowed with significant potential. As a possible future direction for EndoNet, we envision its integration into the QuPath program and its deployment in laboratory testing.

\section*{Funding}
This work was supported by the Ministry of Sciyence and Higher Education of the Russian Federation, agreement No. 075-15-2022-294 dated 15 April 2022. The section concerning collecting and processing samples was supported by the Ministry of Health of the Russian Federation within the framework of State Assignment №121032500100-3. Work of Vishnyakova P. was supported by Russian Science Foundation [grant number 22-75-00048].

\section*{Data availability}
Data will be made available on request. Weights and code for models are provided in paper's repository\cite{github_isp}.

\bibliographystyle{unsrt}  
\bibliography{refs}

\end{document}